\documentclass{PoS}

\title{Lattice QCD calculations of the quark and gluon contributions to the proton spin}

\ShortTitle{Proton spin from lattice QCD}

\author{\speaker{Rajan Gupta}\thanks{Work done as part of the 
    PNDME Collaboration whose other members are Tanmoy Bhattacharya, Vincenzo
    Cirigliano, Yong-Chull Jang, Huey-Wen Lin and Boram Yoon}
  \\ Theoretical Division T-2, Los Alamos National Laboratory, Los
  Alamos, NM 87545, USA\\ E-mail: \email{rg@lanl.gov}}

\abstract{A review of the calculations of the proton's spin using
  lattice QCD is presented.  Results for the three contributions, the
  quark contribution $\sum_{q=u,d,s,c} (\frac{1}{2} {\Delta q})$, the
  total angular momentum of the quarks $J_q$ and of the gluons $J_g$,
  and the orbital angular momentum of the quarks are discussed.  The
  best measured is the the quark contribution $\sum_{q=u,d,s,c}
  (\frac{1}{2} {\Delta q})$, and its analysis is used to discuss the
  relative merits of calculations by the PNDME, ETMC and $\chi$QCD
  collaborations and the level of control over systematic errors
  achieved in each.  The result by the PNMDE collaboration,
  $\sum_{q=u,d,s} \left[ \frac{1}{2} {\Delta q} \right] =
  0.143(31)(36) $, is consistent with the COMPASS analysis $0.13 <
  \frac{1}{2} \Delta \Sigma < 0.18$. Results for $J_q$ and $J_g$
  by the ETMC collaborations are also consistent with
  phenomenology. Lastly, I review first results from the LHPC
  collaboration for the calculation of the orbital angular momentum of
  the quarks. With much larger computing resources anticipated over
  the next five years, high precision results for all three will
  become available and provide a detailed description of their
  relative contributions to the nucleon spin. }

\FullConference{23rd International Spin Physics Symposium - SPIN2018 -\\
		10-14 September, 2018\\
		Ferrara, Italy}

\begin{document}

\section{Introduction}
\label{sec:intro}

The spin is a fundamental defining property of the proton along with
its mass, charge and magnetic moment.  The simplest quark model
picture would indicate that $S=1/2$ arises as a vector sum of the
spins of the three valence quarks.  In 1987, the European Muon
Collaboration presented the remarkable result that the sum of the
spins of the quarks contributes less than half of the total spin of
the proton based on measurements of the spin asymmetry in polarized
deep inelastic scattering~\cite{Ashman:1987hv}. This unexpected result
was termed the ``proton spin crisis''.  The recent result of the
COMPASS analysis is that the intrisic quark contribution to the
proton's spin is only about 30\%, $0.13 < \frac{1}{2} \Delta \Sigma <
0.18$ at 3~GeV${}^2$~\cite{Adolph:2015saz}.

Theoretically, the spin of the proton can be obtained by measuring a
set of matrix elements of operators composed of quarks and gluons
within the ground state of the nucleon. In this review, I will work
with the gauge invariant decomposition of the nucleon's total spin
proposed by Ji~\cite{Ji:1996ek}
\begin{equation}
\frac{1}{2} =  \sum_{q=u,d,s,c,\cdot} \left(\frac{1}{2} {\Delta q} + L_q \right) + J_g 
\label{eq:Ji}
\end{equation}
where ${\Delta q} \equiv {\Delta \Sigma_q} \equiv \langle 1
\rangle_{\Delta q^+} \equiv g_A^q$ is the contribution of the
intrinsic spin of a quark with flavor $q$; $L_q$ is the orbital
angular momentum of that quark; and $J_g$ is the total angular
momentum of the gluons.  Thus, to explain the spin of the proton
starting from QCD, one needs to calculate the contributions of all
three terms. Of the three, the best determined is the first term,
$\frac{1}{2} \Delta \Sigma \equiv \sum_{q=u,d,s} \frac{1}{2} {\Delta
  q} $. Results for which are presented here and have also been reviewed in
the recent FLAG 2019 report~\cite{Aoki:2019cca}.  \looseness-1

Lattice QCD can unravel the mystery of where the proton gets its spin
by measuring, from first principles, the matrix elements of
appropriate quark and gluon operators within the nucleon state.

\begin{figure}[tpb] 
\centerline{
\includegraphics[width=0.32\linewidth]{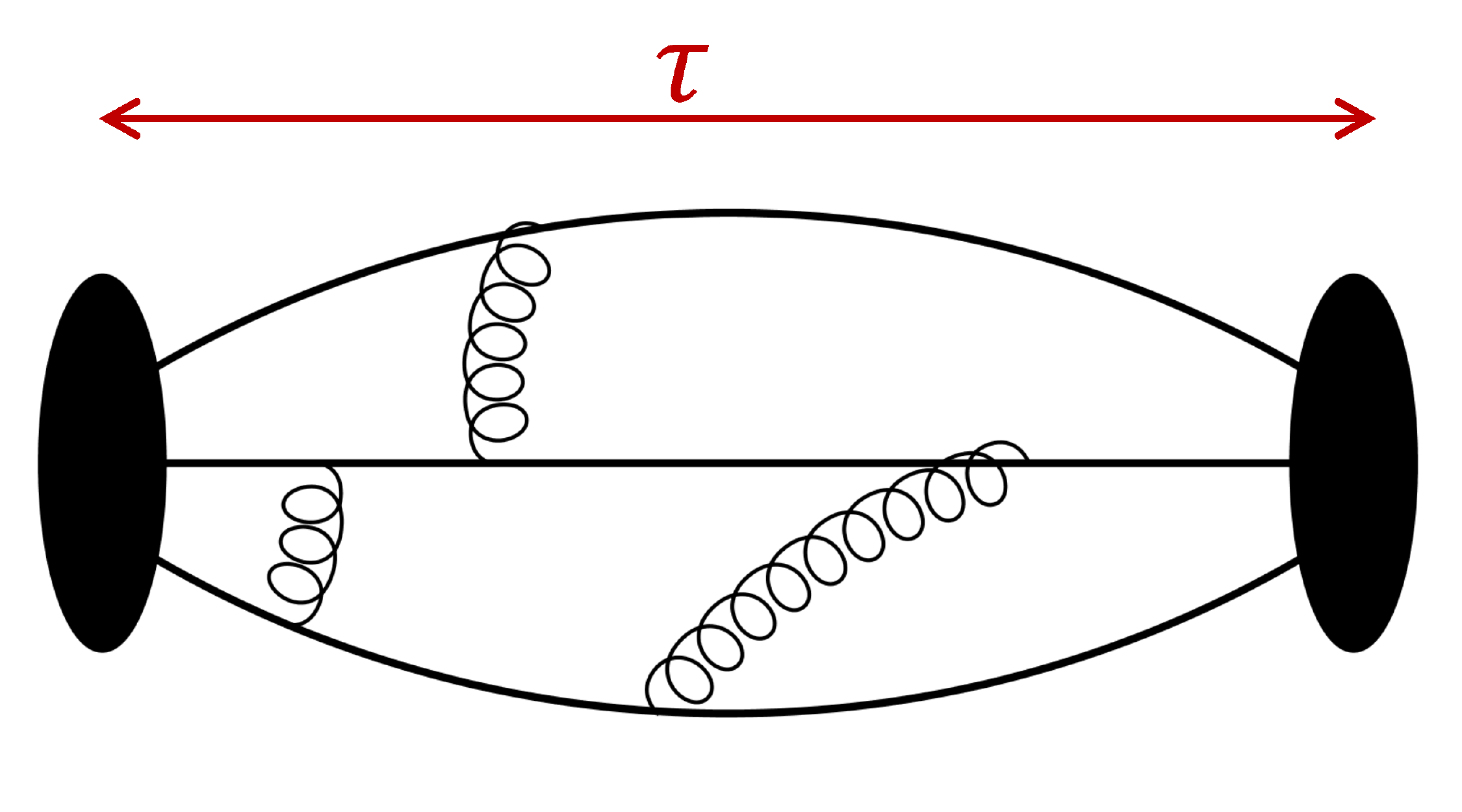} 
\includegraphics[width=0.32\linewidth]{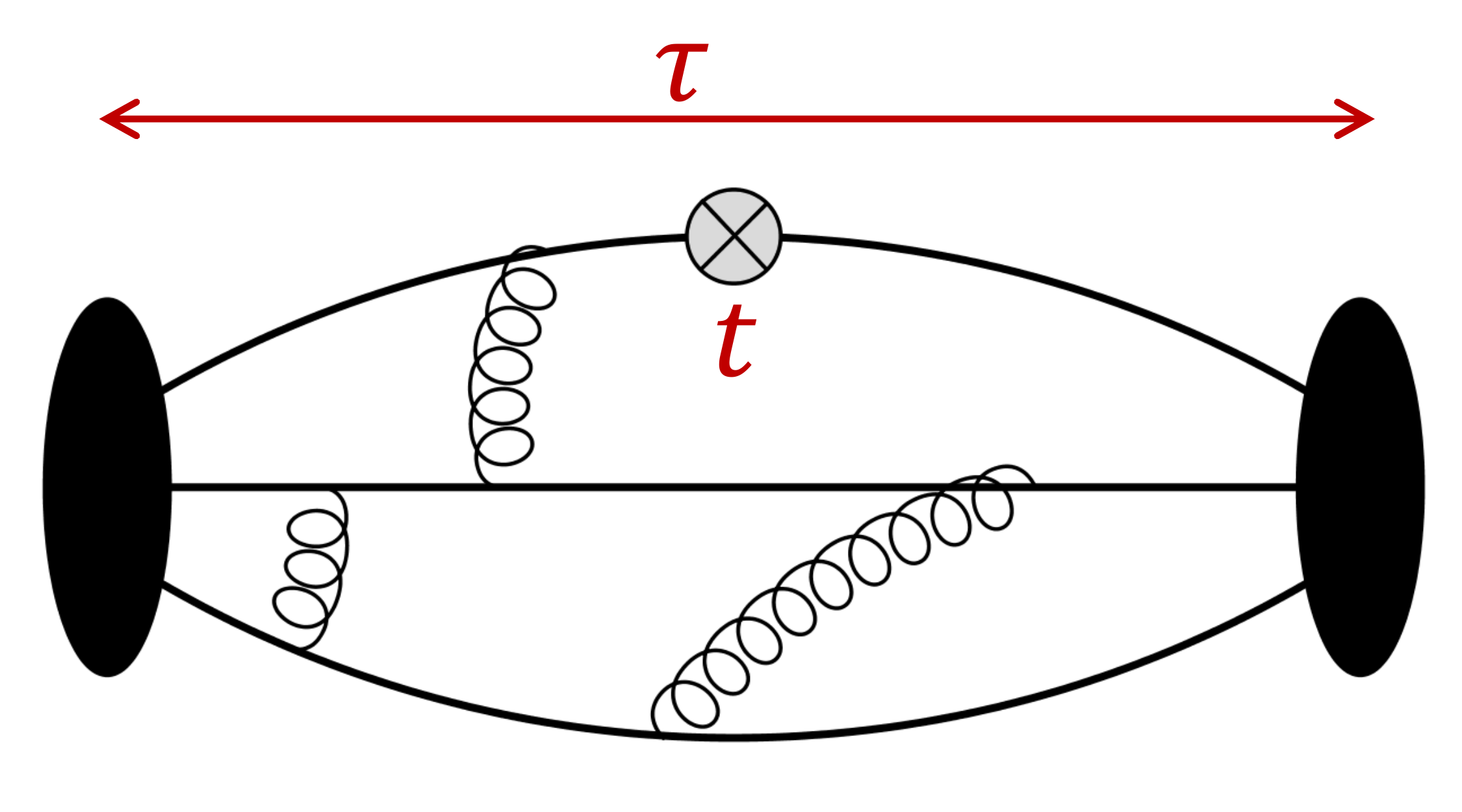}
\includegraphics[width=0.32\linewidth]{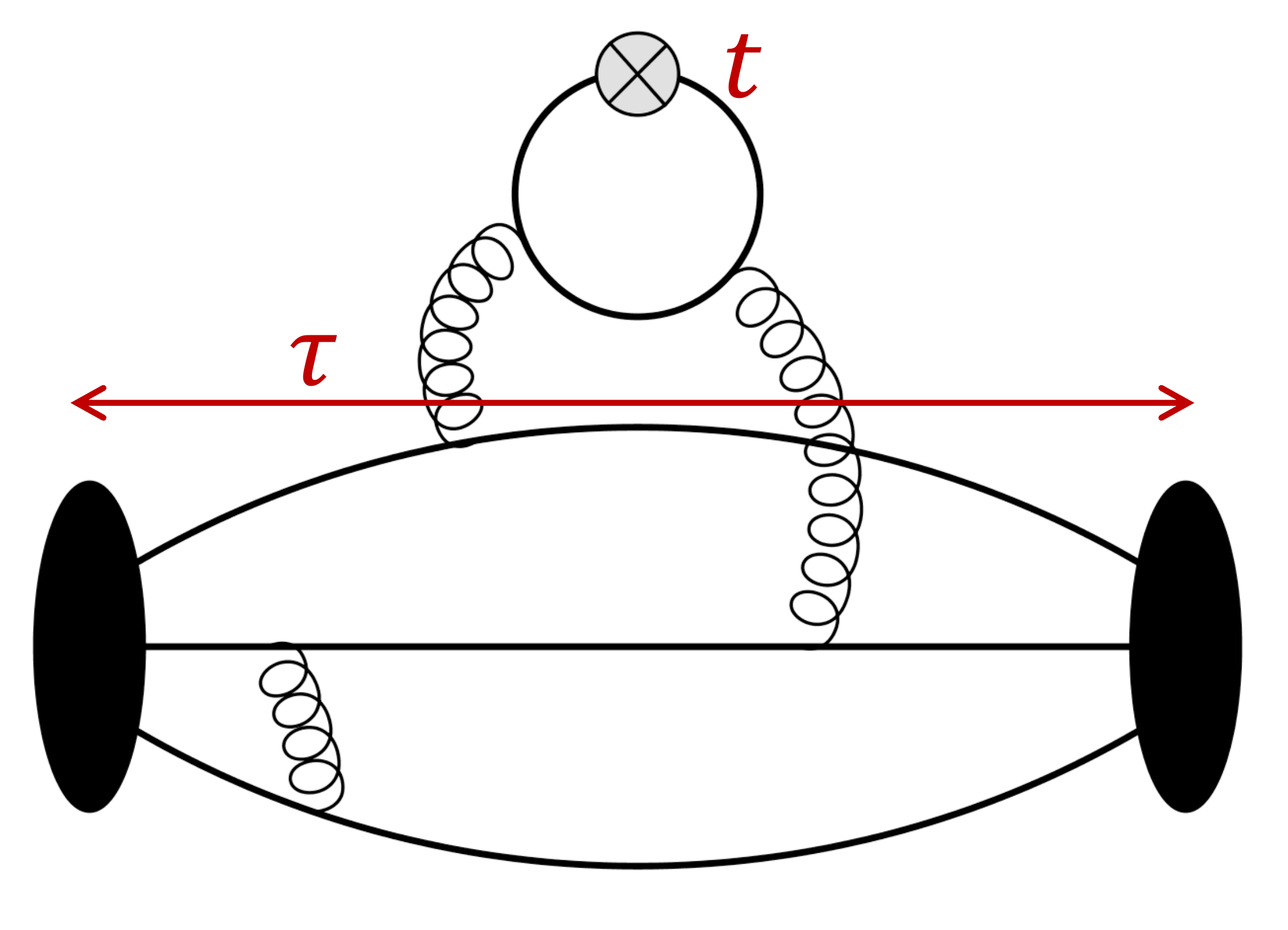}
}
\caption{Illustration of the two- and three-point correlation
  functions calculated to extract the ground state nucleon matrix
  elements. (Left) the nucleon two-point function. (Middle) the
  connected three-point function with source-sink separation $\tau$
  and operator insertion time slice $t$. (Right) the analogue
  disconnected three-point function that contributes to the flavor diagonal operators. }
\label{fig:con_disc}
\end{figure}

As orientation, the connection of the lattice QCD calculation of
correlation functions from which matrix elements are extracted to an
introductory course in non-relativistic quantum mechanics is
conceptually simple. Simulations of lattice QCD using the path
integral representation of the quantum field theory, provide the full
relativistic Fock space wavefunction of a state (mesons or baryons)
within which matrix elements can calculated.  An illustration of the
2- and 3-point functions with the source and sink separated by
Euclidean time $\tau$ and with the insertion of a quark bilinear
operator $\overline{q} \Gamma q$, with $\Gamma$ one of the sixteen
Dirac matrices, is given in Fig.~\ref{fig:con_disc}. The three points
are the source and sink at which the nucleon is created and
annihilated using a suitable interpolating operator, and the timeslice
of insertion on a quark line (valence or sea) of the quark bilinear operator whose
matrix element is desired. Evaluating such correlation functions on
each configuration, represents a ``path''.  The full nonperturbative
wavefunction, and the matrix element of an operator within it, is then
built up by the sum over all the ``paths'', with the contribution of a
given configuration weighted by its action.  Expectations values
of these correlation functions are finally obtained as the ensemble
average over gauge configurations. To summarize, the
elaborate machinery of lattice QCD, summarized briefly in
Sec.~\ref{sec:flowchart}, provides the non-perturbative wavefunction
of the nucleon state within which the matrix elements of various
operators can be calculated as ensemble averages.

To get high precision estimates for such nucleon n-point functions,
the phase space of the path integral has to adequately covered, i.e.,
the ensemble should consist of a sufficiently large number of
decorrelated gauge configurations. Next, all the systematic
uncertainties introduced by discretizing QCD on a 4-d lattice need to
be understood and controlled. Finally, lattice results that can be
compared with experiments or phenomenology are obtained by performing
a chiral-continuum-finite-volume (CCFV) fit to the data obtained over
a range of values of $a$, $M_\pi$ and $L$ and evaluating the result at
the physical pion mass $M_\pi = 135$~MeV, taking the continuum limit defined by
the lattice spacing $a \to 0$ and extrapolating in lattice size $L \to
\infty$. A priori, one does not know how large a given systematic
uncertainty in a given quantity is. It is largely determined a
posteriori from the CCFV fits to the data. To increase the reliability
of the CCFV fits and to control the total error requires high
statistics data.  In this review, I will devote considerable attention
to how well the statistical and various systematic errors have been
controlled and estimated in the various calculations.

%
%

\section{Flowchart of lattice QCD Calculation}
\label{sec:flowchart}

The lattice methodology for the calculation of the contribution of the
intrinsic spin of the quarks to the proton spin is mature. I will use it to exemplify, 
very briefly, the steps in the calculation. 
\begin{itemize}
\item
Formulate QCD on a finite 4-D Euclidean grid with lattice spacing
$a$. This step defines the action ${\cal A} = {\cal {A}}_G + \sum_i \overline q D_i q$ for the gauge and the
quarks fields, and introduces discretization and finite volume
errors. Here $D_i$ is the Dirac action for flavor $i$ and the sum is over the quark flavors.
Since there is no one perfect lattice action that preserves
all the properties of continuum QCD at finite $a$, a number of
different actions have been used in simulations. They typically differ
in how well the continuum chiral symmetry is preserved, the order
$O(a^n)$ of the discretization errors, and the cost of generation of
ensembles of gauge configurations.
\item
In the path integral formulation of quantum field theory used in
numerical simulations, the quark degrees of freedom are integrated
out. The resulting Boltzmann weight, ${\cal{A}}={\cal{A}}_G + \sum_i {\rm Tr} \log
D_i$, which is used to generate ensembles of background gauge fields,
becomes a functional only of the gauge fields. The effects of quarks on the QCD vacuum 
are included through the term $\sum_i {\rm Tr} \log D_i$ in the Boltzmann weight used to generate lattices. 
\item
A suite of ensembles of gauge configurations at multiple values of the
lattice spacing and the light quark mass are generated with the chosen
discretized action $\cal A$ using a Markov Chain Monte Carlo method
with importance sampling. Simulations at a range of light quark masses
at a fixed value of $a$ (equivalently the gauge coupling) are carried
out to understand the chiral behavior of the observable and to improve
the reliability of the extraction at the physical value taken to be $M_{\pi^0} =
135$~MeV.  With improvements in both algorithms and the computing
power, current simulations include points at/near the physical pion
mass. This has greatly improved the chiral fits and the extraction of
results at the physical point $M_\pi = 135$~MeV. 
\item
Most calculations
have, so far, been done assuming isospin symmetry, $m_u= m_d$, i.e.,
two degenerate light quark flavors. Thus effects proportional to $m_d-m_u$ are 
neglected. These are expected to be small for the quantities discussed here.
\item
Since the strange and charm quarks are relatively heavy, their
contribution to the non-perturbative vacuum are included in the
generation of gauge configurations with masses tuned to their physical
values using appropriate spectral quantities, for example the masses
of the $\Omega$ baryon and the $\eta_c$ meson. Thus, no fits in these
quark masses are needed to get physical estimates. Simulations including 
these flavors are labeled 2+1 and 2+1+1 flavor calculations,
respectively. \looseness-1
\item
The basic building blocks of the correlation functions are the gauge
links and the quark propagators. Quark propagators, given by the
inverse of the Dirac operator on a given configuration, are computed using Krylov solvers.
Inverting the Dirac matrix, whether in gauge configuration generation or for
quark propagators, is computationally the most expensive
part of the calculation. The current algorithm of choice is the
algebric Multigrid.  
\item
Since the generation of gauge configurations is
expensive, multiple measurements of correlation functions are made on
each gauge configurations to increase the statistics. This exploits
the fact that a large volume lattice can be considered to consist of
many essentially decorrelated subvolumes.
\item
A large set of gauge invariant correlation functions (for example, for
extracting matrix elements of different operators) are calculated at
the same time by contracting the spin and color indices of quark
propagators and gauge links in all possible combinations.
\item
Expectation values are constructed by averaging these correlation
functions over measurements, i.e., over both multiple source points on a
given configuration and over the ensemble of gauge configurations.
\item
Observables $O$, such as masses and matrix elements, are extracted
from these expectation values using the spectral decomposition of the
correlation functions.  In this decomposition, the spectrum in a
finite volume is defined by the eigenvalues of the transfer matrix.
\item
Different versions of an operator can be defined on the lattice and
one can use any of these to calculate a given matrix elements. At
finite $a$, results for different bare operators will, in general,
differ. Renormalizing the operators removes the variation, and their
matrix elements are finite in the continuum limit. In addition, to
connect renormalized lattice results to those used by
phenomenologists, the renormalization process includes a
multiplicative matching factor from the lattice to some continuum
scheme such as $\overline{MS}$ at a given scale, typically taken to be
$\mu =2$~GeV.
\item
Lattice results obtained using renormalized operators depend on the
lattice spacing, the pion mass (surrogate for the light quark mass),
and the lattice size $L$. To obtain their physical value, $O_{\rm
  ph}$, lattice artifacts are removed by extrapolating $O(a, M_\pi,
L)$ to $a \to 0$, $M_\pi=135$~MeV and $L \to \infty$ using fits to
data at multiple values of $a$, $M_\pi$ and $L$.  These combined
chiral-continuum-finite-volume (CCFV) fits are made using ansatz that are observable
specific and physically motivated. For example, chiral perturbation
theory is used to deduce the form of the correction terms with respect to $M_\pi$. The
CCFV ansatz, with just the leader order correction terms, that is
commonly used to fit the lattice data for the axial charges is
\begin{equation}
  g_{A}^{u,d,s} (a,M_\pi,L) = g_A^{u,d,s}|_{\rm ph} + c_2 a + c_3 M_\pi^2 + c_4 M_\pi^2 e^{-M_\pi L} \,,
\label{eq:CCFV} 
\end{equation}
for lattice formulations in which discretization errors begin at $O(a)$ (PNDME). For 
the $\chi$QCD and ETMC calculations, this term should be read as $c_2 a^2$. 

\end{itemize}

\subsection{Excited-State Contamination}
\label{sec:ESC}

\begin{figure}[tpb] 
\centerline{
\includegraphics[width=0.47\linewidth]{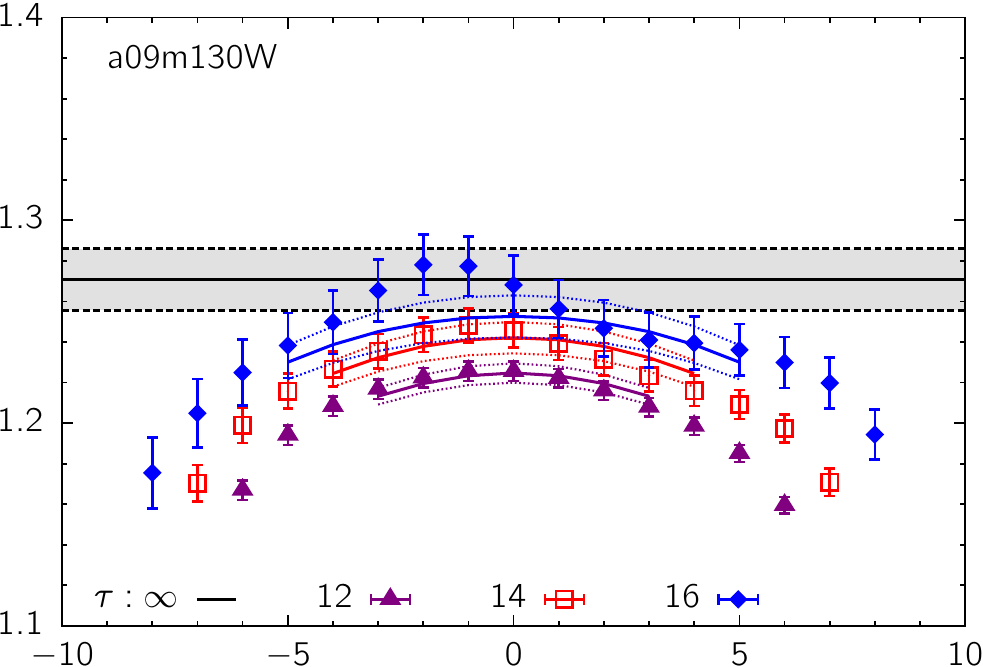} 
\includegraphics[width=0.47\linewidth]{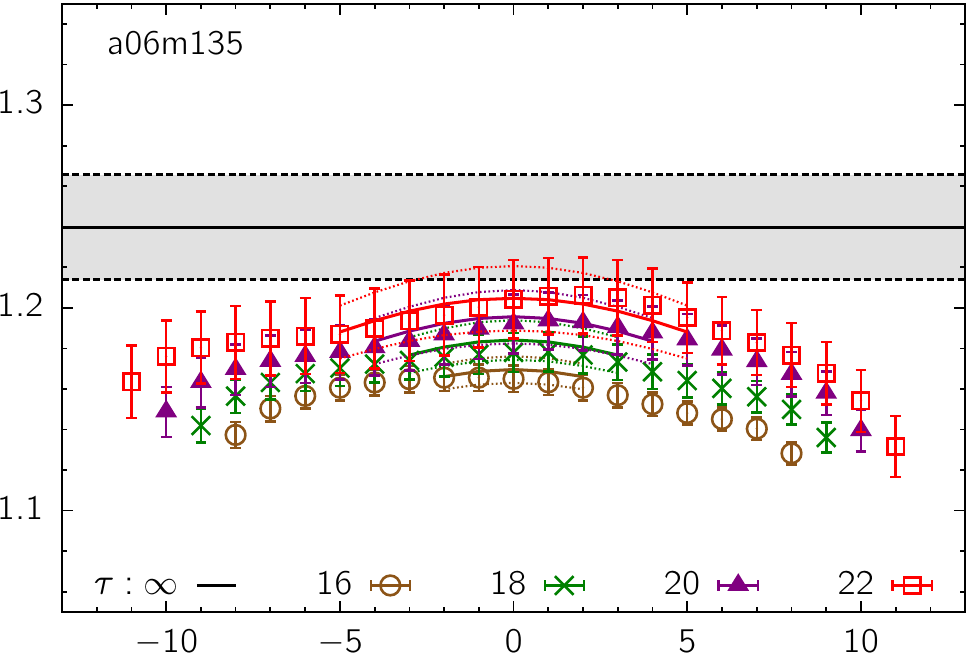}}
\caption{Illustration of the excited-state contamination in the
  extraction of $g_A^{u-d}$. Data from the two $M_\pi \approx 135$~MeV
  ensembles analyzed by the PNDME collaboration~\protect\cite{Gupta:2018qil} are shown as a
  function of $t-\tau/2$ for various source-sink separations $\tau$. The grey
  band is the result for $g_A^{u-d}$ obtained in the $\tau \to \infty$
  limit from a $3^\ast$-state fit. The colored lines show the fit for
  various values of $\tau$. }
\label{fig:ESC}
\end{figure}

An additional systematic error particularly relevant to the
calculations of matrix elements within nucleon states is excited-state
contamination. This is because interpolating operators used to create
and annihilate nucleon states at either end the n-point correlation functions
couple not just to the ground state nucleon but to all excitations and
multiparticle states with the same quantum numbers.  For baryons, the
contribution of excited states is observed to be large because the
number of states between 1.2--1.5 GeV grows with lattice volume.
Their contributions need to be removed for each observable and on each ensemble before 
CCFV fits are made to get $O^{\rm ph}$. Two examples of excited-state
contamination in the extraction of $g_A^{u-d}$ and its control by the
PNDME Collaboration~\cite{Gupta:2018qil} using fits with up to three
states in the spectral decomposition are shown in Fig.~\ref{fig:ESC}.

In the calculations reviewed, the excited state and the chiral-continuum fits were
done separately for the connected and disconnected contributions. This is because 
the ranges of source-sink separation studied are typically different as are the 
number of ensembles analyzed. Such a separate analysis introduces an additional 
systematic that is judged to be small as explained
in Ref.~\cite{Lin:2018obj}.

To obtain high-precision lattice results requires control over both
statistical and the various systematic errors, i.e., excited-state
contamination and those removed by the CCFV fits.  In
Section~\ref{sec:errors}, I provide a critical analysis of the
strengths and limitations of three calculations,
PNDME~\cite{Lin:2018obj}, ETMC~\cite{Alexandrou:2017oeh} and
$\chi$QCD~\cite{Liang:2018pis}, of the quark contribution to the
proton spin.

%
%

\section{Intrinsic quark contribution to the proton spin}
\label{sec:errors}

The intrinsic quark contribution to the proton spin is given by the flavor diagonal 
axial charges $g_A^q$. These are given by 
the matrix element of the flavor diagonal axial current, $\overline{q} \gamma_\mu
\gamma_5 q$, 
\begin{equation}
g_A^q \overline{u}_N \gamma_\mu \gamma_5 u_N \!=\!  \langle N| Z_A \overline{q} \gamma_\mu \gamma_5 q | N \rangle 
\label{eq:gAdef}
\end{equation}
where $Z_A$ is the renormalization constant and $u_N$ is the neutron spinor. 
In addition to quantifying the contribution of the quarks to the nucleon
spin, 
\begin{equation}
g_A^q \equiv \Delta q \!=\! \int_0^1 dx (\Delta q(x) + \Delta \overline{q} (x) )
\end{equation}
is also the first Mellin moment of the polarized parton distribution
function (PDF) integrated over the momentum fraction
$x$~\cite{Lin:2017snn}. Thse are measured in semi-inclusive deep
inelastic scattering experiments.  The charges, $g_A^{u,d,s}$, also
quantify the strength of the spin-dependent interaction of dark matter
with nucleons~\cite{Fitzpatrick:2012ib,Hill:2014yxa}.  Of these,
$\Delta s$ is the least well known and current phenomenological
analyses~\cite{Lin:2017snn} often rely on assumptions such as SU(3)
symmetry and $\Delta s = \Delta \overline{s}$.

The most costly part of the calculation of $g_A^{u,d,s}$ is the
contribution due to disconnected quark loops illustrated in
Fig~\ref{fig:con_disc} (right). It is computed stochastically and then
correlated with the nucleon two-point function. The resulting
three-point function is then averaged over the ensemble of gauge
configurations. The statistical error comes from both the stochastic
evaluation (on each configuration) of the quark loop and it's
correlation with the nucleon 2-point function, and the ensemble
average of the three-point function.  Since the computational cost
increases significantly as $M_\pi \to 135$~MeV, calculations of
$g_A^{u,d,s}$ at the physical pion mass have started to be done only
recently. The three results discussed in the next section include both
disconnected contributions and evaluation of $g_A^q$ at $M_\pi = 135$~MeV.

\section{Lattice calculations of $g_A^{u,d,s}$}
\label{sec:gA}

An overview of the lattice parameters of the results from three
collaborations, PNDME~\cite{Lin:2018obj},
ETMC~\cite{Alexandrou:2017oeh} and $\chi$QCD~\cite{Liang:2018pis} that
have presented results at the physical pion mass are given in
Table~\ref{tab:lattice}.  I will not review the work by
JLQCD~\cite{Yamanaka:2018uud}, LHPC~\cite{Green:2017keo} and
Engelhardt~\cite{Engelhardt:2012gd} as they were done either at a
single lattice spacing and/or with heavy quarks, and therefore cannot 
be compared to experimental/phenomenological results.

The results for connected and disconnected contributions for the three
calculations reviewed are summarized in Table~\ref{tab:gACD}, and the final
results in Table~\ref{tab:gA}.

\begin{table}[b]
\setlength{\tabcolsep}{6pt}
\begin{tabular}{|l|l|l|c|c|c|c|}
\hline
Collaboration                  & $N_f$   & Formulation            &$\#$ of   &  $a$         &  $M_\pi^{\rm val} $         \\
                               &         &                        &Ensembles &  (fm)        &  (MeV)            \\
\hline
PNDME~\cite{Lin:2018obj}       & 2+1+1   & Clover-on-HISQ         &  1       &  $0.15$  &  $321$                \\
                               &         &                        &  4       &  $0.12$  &  $310, 228$           \\
                               &         &                        &  3       &  $0.087$ &  $313, 226, 138$      \\
                               &         &                        &  3       &  $0.057$ &  $320, 235, 136$      \\
\hline
$\chi$QCD~\cite{Liang:2018pis} & 2+1     & Overlap-on-Domain Wall &  1       &  $0.143$ &  147--327             \\
                               &         & (Partially quenched)   &  1       &  $0.11$  &  254--389             \\
                               &         &                        &  1       &  $0.083$ &  260--410             \\
\hline
ETMC~\cite{Alexandrou:2017oeh} & 2       & Twisted mass           &  1       &  0.094   &  130                  \\
\hline
\end{tabular}
\caption{\label{tab:lattice}Lattice parameters of the three
  calculations discussed.  The $\chi$QCD calculation consists of 5 or
  6 values of the valence quark mass on each of the 3 ensembles constituting a partially quenched calculation. 
}
\end{table}

\begin{table}[b]
\setlength{\tabcolsep}{6pt}
\begin{tabular}{|l|c|c|c|c|c|}
\hline
Collaboration                  & $g_A^{u}|_{\rm Conn}$ &  $g_A^d|_{\rm Conn}$      &  $g_A^{u,d}|_{\rm disc}$ &  $g_A^{s}|_{\rm disc}$   \\
\hline
PNDME~\cite{Lin:2018obj}       & 0.895(21)             &  -0.320(12)               &  -0.118(14)        &  -0.053(8)                            \\
$\chi$QCD~\cite{Liang:2018pis} & 0.917(13(28)          &  -0.337(10)(10)           &  -0.070(12)(15)    &  -0.035(6)(7)                         \\
ETMC~\cite{Alexandrou:2017oeh} & 0.904(40)             &  -0.305(28)               &  -0.075(14)        &  -0.042(10)(2)                        \\
\hline
\end{tabular}
\caption{\label{tab:gACD}Results for the flavor diagonal axial charges
  $g_A^{u,d,s} = {\Delta q}$ for the proton.  Results for the neutron
  can be obtained by interchanging the $u $ and $d$ flavor
  indices. All lattice results are quoted in $\overline{MS}$ scheme at
  2~GeV${}^2$.  }
\end{table}

\begin{table}[b]
\setlength{\tabcolsep}{6pt}
\begin{tabular}{|l|c|c|c|c|c|}
\hline
Collaboration                  & $g_A^{u-d}$      &  $g_A^u$      &  $g_A^d$         &  $g_A^s$       &  $\sum_{q=u,d,s} (\frac{1}{2} {\Delta q})$  \\
\hline
PNDME~\cite{Lin:2018obj}       & 1.218(25)(30)   &  0.777(25)(30) &  -0.438(18)(34)  &  -0.053(8)     &  0.143(31)(36)                              \\
$\chi$QCD~\cite{Liang:2018pis} & 1.254(16)(30)   &  0.847(18)(32) &  -0.407(16)(18)  &  -0.035(6)(7)  &  0.203(13)(19)                              \\
ETMC~\cite{Alexandrou:2017oeh} & 1.212(40)       &  0.830(26)(4)  &  -0.386(16)(6)   &  -0.042(10)(2) &  0.201(17)(5)                               \\
\hline
\end{tabular}
\caption{\label{tab:gA}Results for the flavor diagonal axial charges
  $g_A^{u,d,s} = {\Delta q}$ for the proton.  Results for the neutron
  can be obtained by interchanging the $u $ and $d$ flavor indices.  }
\end{table}

There are two obvious questions looking at the results in
Tables~\ref{tab:gACD} and~\ref{tab:gA}: are the statistical and
systematic errors in the three calculations equally well understood
and controlled, and how much of the difference between the final
results is due to the difference in the lattice parameters, listed in
Table~\ref{tab:lattice}, that define the three calculations. Before
answering the questions, I summarize the strengths and limitations of
the three calculations to highlight the differences.

\subsection{PNDME calculation}
\label{sec:errorsPNDME}

The connected parts of the PNDME~18A~\cite{Lin:2018obj} results were
obtained using eleven 2+1+1 flavour HISQ ensembles generated by the
the MILC collaboration with $a \approx 0.057$, 0.87, 0.12 and 0.15~fm;
$ M_\pi \approx 135$, 220 and 320~MeV; and $3.3 < M_\pi L < 5.5$. The
light disconnected contributions were obtained on six of these
ensembles with the lowest pion mass $M_\pi \approx 220$~MeV, while the
strange disconnected contributions were obtained on seven ensembles,
i.e., including an additional one at $a \approx 0.087$~fm and $M_\pi
\approx 135$~MeV. The CCFV fits to the connected contribution were
done using the ansatz given in Eq.~(\ref{eq:CCFV}), and the finite
volume correction was dropped for the analysis of the disconnected
data.

The strengths of the PNDME calculation~\cite{Lin:2018obj} with $2+1+1$ flavors of dynamical quarks are:
\begin{itemize}
\item
High statistical precision with $O(10^5)$ measurements performed on each ensemble. 
\item
The data on the 11 ensembles cover a reasonable range in all three variables: 
$ 0.057 < a < 0.15$~fm, $135 < M_\pi < 320$~MeV and $3.3 < M_\pi L < 5.5$.
\item
The analysis of the excited-state contamination, discussed in
Sec.~\ref{sec:flowchart}, was done using three-state fits for the
connected contribution and two-state fits for the disconnected
contributions. Data at 4--5 values of the source sink separation
$\tau$ in the range 1--1.5~fm were used in these fits.
\item
The CCFV fit was carried out keeping the leading terms in $a$,
$M_\pi^2$ and $M_\pi L$ as defined in Eq.~(\ref{eq:CCFV}).  Data from the two physical pion 
mass ensembles anchored the fit versus $M_\pi$. The Akaike Information
Criteria~\cite{1100705} was used to justify not including higher order
corrections, otherwise the fits would be over-parameterized.
\item
The CCFV fits are done separately for both the connected and disconnected contributions. 
The dominant variation in both was shown to be versus $a$.
\end{itemize}
The limitations of the PNDME calculation are:
\begin{itemize}
\item
The mixed action, clover-on-HISQ, formulation is expected to give results for QCD in the $a \to 0$ limit. 
Ultimately, a confirmation using a unitary formulation is needed. 
\item
The renormalization of the flavor diagonal charges is done assuming
$Z_A^{\rm singlet} = Z_A^{\rm nonsinglet}$. While this has been
validated to hold to within a percent by the ETMC and $\chi$QCD
calculations, it needs to be confirmed for the clover-on-HISQ
ensembles.
\item
Their estimate of the isovector axial charge $g_A^{u-d} =
1.218(27)(30) $ is about 5\% below the experimental value 1.277(2).
The authors account for this deviation in the second
systematic uncertainty of $0.030$ quoted in both $g_A^u$ and $g_A^d$.
\end{itemize}

\subsection{$\chi$QCD calculation}
\label{sec:errorschiQCD}

The $\chi$QCD~\cite{Liang:2018pis} calculation used three ensembles of
$2+1$ flavors of dynamical domain-wall quarks generated by the
RBC/UKQCD collaboration.  Since two different discretizations of
domain-wall fermions were used, the discretization effects in the CCFV
fits require two separate $O(a^2)$ terms.  The strengths of the
$\chi$QCD~\cite{Liang:2018pis} calculation are:
\begin{itemize}
\item
Both the sea and valence quark actions in the overlap-on-domain-wall formalism 
preserve the continuum chiral symmetry at finite $a$. 
\item
The excited-state contamination is controlled using 2 states in the spectral decomposition of the 3-point data 
obtained at 4--5 values of the source sink separation $\tau$. 
\item
Both renormalization factors, $Z_A^{\rm singlet}$ and $Z_A^{\rm nonsinglet}$, were calculated. 
They were found to agree to within a percent. 
\item
The estimate of $g_A^{u-d} = 1.254(16)(30) $ is consistent with the experimental value. 
\end{itemize}
The limitations of the $\chi$QCD calculation are:
\begin{itemize}
\item
The overlap-on-domain-wall formulation is also non-unitary. 
\item
Only three approximate ``unitary'' points with lattice spacings 0.143,
0.11 and 0.083~fm and pion masses $M_\pi = 171$, 337 and 302~MeV for
the sea quarks, respectively, were analyzed.  At each $a$, partially
quenched data at 4--5 addition pion masses was collected.  All the
points (unitary and partially quenched) were analyzed together. In the
chiral fit to this partially quenched data, possible dependence on
$M_\pi^{\rm sea}$ was neglected and the data were fit versus only $M_\pi^{\rm
  valence}$.
\item
The CCFV fit used two terms of the form 
$c_3 M_\pi^{2,{\rm sea}} + c_3^v M_\pi^{2, {\rm valence}}$
in Eq.~(\ref{eq:CCFV}), however, in practice, it was only sensitive
to $M_\pi^{\rm valence}$.  In the end, with only 3 ``unitary'' data points, Baysian priors were used to stabilize the
two coefficients of the $O(a^2)$ terms and the dependence on $M_\pi^{\rm
  sea}$ (sensitive only to the three approximately unitary points) and
finite lattice size was neglected.
\end{itemize}

\subsection{ETMC calculation}
\label{sec:errorsETMC}

The strengths of the ETMC~\cite{Alexandrou:2017oeh} calculation with $2$ flavors of dynamical quarks are:
\begin{itemize}
\item
This is a unitary calculation. The same action, twisted mass with a
clover term, is used for both the sea and valence quarks.
\item
The discretization errors in the twisted mass with a clover term formalism 
start at $O(a^2)$. 
\item
No extrapolation in $M_\pi$ was needed.
\item
Both renormalization factors, $Z_A^{\rm singlet}$ and $Z_A^{\rm
  nonsinglet}$, were calculated and found to agree to within a
percent.
\end{itemize}
Limitations of the ETMC calculation are:
\begin{itemize}
\item
The calculation used a single ensemble with $M_\pi=130$~MeV, $a=0.094$
and a relatively small $M_\pi L = 2.98$.  Thus discretization errors
and finite lattice size corrections cannot be assessed. 
\item
The estimate of $g_A^{u-d} = 1.212(40) $ is about 5\% below the experimental value 1.277(2).
\end{itemize}

\subsection{My overall assessment of $g_A^{u,d,s}\equiv  {\Delta \Sigma_q}$}
\label{sec:bottomline}

\begin{figure}[tpb] 
\centerline{
\includegraphics[width=0.47\linewidth]{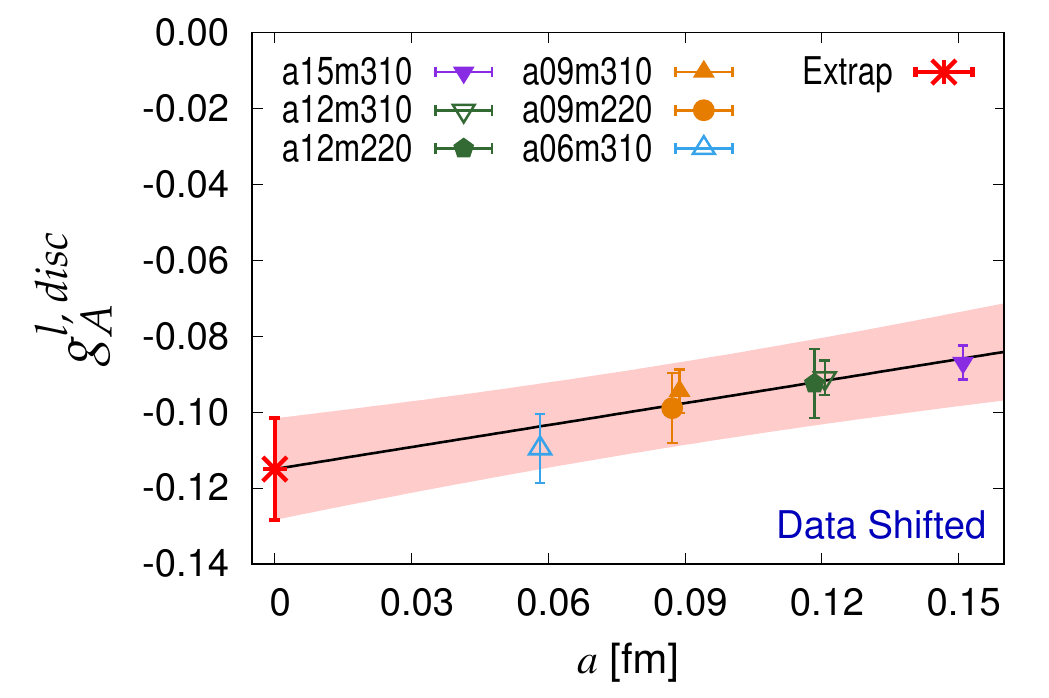}
\includegraphics[width=0.47\linewidth]{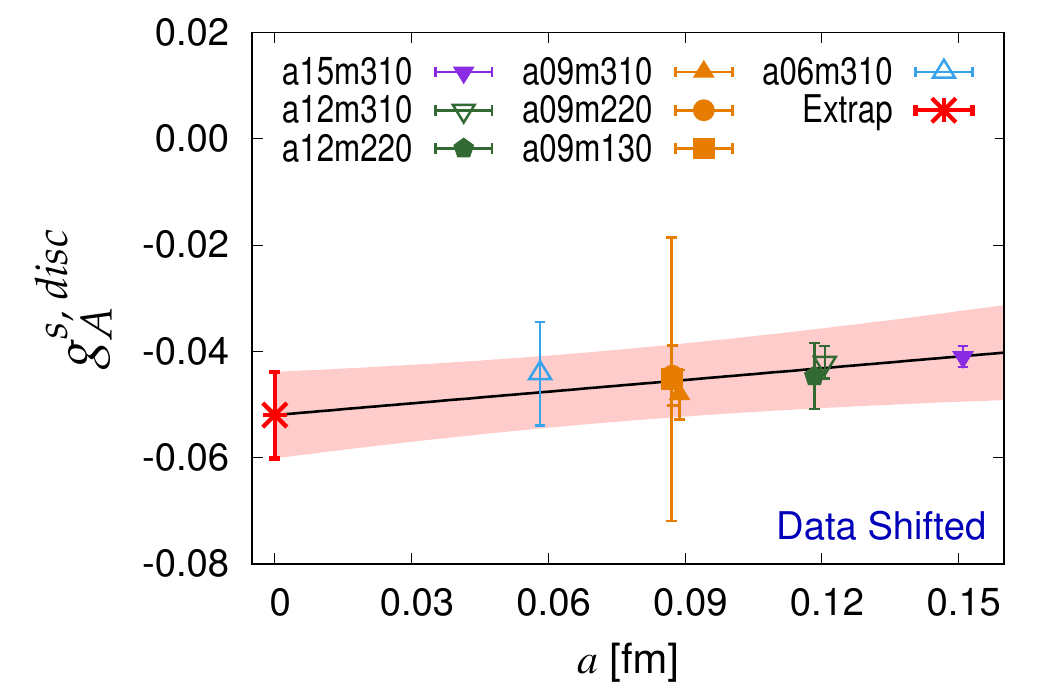}}
\caption{The chiral-continuum extrapolation of the renormalized
  $g_A^{l,{\rm disc}}$ and $g_A^{s,{\rm disc}}$ data using the ansatz
  given in Eq.~(\protect\ref{eq:CCFV}).  The pink band shows the
  result of the simultaneous fit plotted versus $a$. The data points
  have been shifted by extrapolating them to the physical point
  $M_\pi=135$~MeV using the fit.  }
\label{fig:extrap}
\end{figure}

Given that the three calculations differ in almost all aspects, it
would seem unlikely that a simple explanation for the difference in
the results shown in Table~\ref{tab:gA} can be presented.  It turns
out that the observed difference can be explained by the $a$
dependence found in the PNDME CCFV fits for the disconnected contributions 
shown in Fig.~\ref{fig:extrap} if one assumes that there is
no significant dependence on $N_f$, the lattice actions and the
lattice size, and the excited-state fits and the chiral extrapolation are
equally reliable. 

%


Fig.~\ref{fig:extrap} shows that the change between $a\approx 0.09$~fm
and $a=0$ was found to be $\approx -0.04$ for $g_A^{l,{\rm disc}}$ and
$\approx -0.01$ for $g_A^{s,{\rm disc}}$.  Assuming that the same
pattern of discretization corrections is applicable to the $\chi$QCD
and ETMC results, then their values for $g_A^u$ would be smaller by
$0.04$, those for $g_A^d$ more negative by $0.04$, and those for
$g_A^s$ more negative by $0.01$.  With these corrections, the results
for the individual $g_A^{u,d,s}$ and for the sum $\frac{1}{2} {\Delta
  \Sigma}$ from the three calculations would overlap. Future higher
precision data from more ensembles is, of course, necessary to validate
this simple explanation.

\section{Total angular momentum of quarks and gluons}
\label{sec:quarks}

The total angular momentum operator can be written in  terms of the energy momentum tensor in a 
gauge invariant way as~\cite{Ji:1996ek}
\begin{equation}
J^i = \frac{\epsilon^{ijk}}{2} \int d^3x (T^{0j}x^k - T^{0k}x^j)
\end{equation}
This can be further decomposed in terms of the contribution of quarks, 
\begin{equation}
\vec{J}_q  =  \int d^3x  \overline{q} \left[\vec \gamma \gamma_5 + \vec x \times (-i \vec D) \right] q, 
\end{equation}
and gluons 
\begin{equation}
\vec{J}_g =  \int d^3x (\vec x \times (\vec E \times \vec B) \,.
\end{equation}
To calculate these two contributions on the lattice, one evaluates the matrix elements of 
the following two operators within nucleon states: 
\begin{equation}
{O}_q^{\mu\nu} =  \frac{1}{2} \left[ \overline{q} \gamma^{(\mu} \stackrel{\rightarrow}{ D}{}^{\nu )} q  + \overline{q} \gamma^{(\mu} \stackrel{\leftarrow}{ D}{}^{\nu )} q \right] \,,
\end{equation}
and 
\begin{equation}
{O}_g^{\mu\nu} =  \frac{1}{4} g^{\mu\nu} G^2 - G^{\mu \alpha}G^{\nu}_\alpha  \,.
\end{equation}
The matrix elements of these operators at momentum transfer $Q^2 \equiv (p^\prime -p)^2$ are then decomposed in terms of Lorentz covariant form factors as 
\begin{equation}
\langle N(p^\prime, s^\prime) | {O}_{q,g}^{\mu\nu} | N(p,s) \rangle =  \overline{u}_N (p^\prime, s^\prime) \Lambda_{q,g}^{\mu \nu} u_N(p,s) 
\end{equation}
where $P \equiv (p^\prime +p)/2$,  $u_N$ is the nucleon spinor and 
\begin{equation}
 \Lambda_{q,g}^{\mu \nu} = A_{q,g}(Q^2)  \gamma^{\{ \mu }P{}^{\nu \} } + B_{q,g}(Q^2) \frac{P^{\{ \mu } \sigma{}^{\nu \} \alpha} Q_\alpha}{2M_N} +  C_{q,g}(Q^2) \frac{Q^{\{ \mu} Q^{ \nu \} }}{M_N}
\end{equation}
Here, $M_N$ is the nucleon mass and the curly braces denote that the
two indices within them have to be symmetrized and the traceless part
taken.  From these, the total angular momentum is obtained from the
following combination of the form factors
\begin{equation}
\vec{J}_{q,g}  =  \left[ A_{q,g}(0)  + B_{q,g}(0) \right] \,.
\end{equation}
On the lattice, $A_{q,g}(0)$ can be extracted directly while $B_{q,g}(0)$ is obtained by 
extrapolating data at $Q^2 \ne 0$ to $Q^2 = 0$. 

The flowchart for the calculation of the three- point function from
which $\langle N(p^\prime, s^\prime) | {O}_{q,g}^{\mu\nu} | N(p,s)
\rangle$ are extracted is similar to that described in
Sec.~\ref{sec:flowchart}. There are, however, a number of
additional challenges:
\begin{itemize}
\item
${O}_q^{\mu\nu}$ involves 1-link (one derivative) operators, and both connected and disconnected
contributions need to be calculated.  
\item
${O}_g^{\mu\nu}$ is constructed out of Wilson loops. There is only a disconnected contribution with a noisier statistical signal. 
\item
The matrix elements have to be decomposed in terms of form factors. The form factor $B_{q,g}(0)$ can only be evaluated at $Q^2 \ne 0$ and the data extrapolated to $Q^2 = 0$.
\end{itemize}

\subsection{ETMC Calculation of $J_q$ and $J_g$}
\label{sec:ETMCtotal}

As described above, the calculation of $J_q$ and $J_g$ is significantly harder and only the ETMC collaboration 
has presented results. Some details of the calculation are: 
\begin{itemize}
\item
  The renormalization factor for ${O}_q^{\mu\nu}$ (involving one derivative operators) 
  has been calculated non-perturbatively.
\item
The renormalization of ${O}_g^{\mu\nu}$ and its mixing with the quark
singlet operator has only been carried out in 1-loop perturbation
theory. The mixing is found to be a small correction. 
\item
The stout smearing of gauge links in the operators brings the renormalization 
factor and mixing coefficient closer to their tree-level values~\cite{Alexandrou:2016ekb}. 
\item
The disconnected contribution to $B_{q}(0)$ is found to be smaller than the statistical errors in the connected contributions. 
So $J_{s,c} \approx A_{s,c}(0)$ is used and $B_{q}(0)$ is neglected.
\item
The form factor $B_{g}(0)$ is assumed to be zero, so $J_g = A_g(0)$ is used. 
\item
Checks on $ A_{u-d}(0)$ are made by comparing with phenomenological
values of the mean momentum fraction $\langle x \rangle_{u-d} = A_{u-d}(0)$.
\end{itemize}

Their results are 
\begin{equation}
\vec{J}_{u+d+s}  =  0.255(12)(3)|_{\rm conn} + 0.153(60)(47)|_{\rm disc} = 0.408(61)(48) \,.
\end{equation}
and 
\begin{equation}
\vec{J}_g \approx A_g(0) = \frac{1}{2}\langle x \rangle_g  =  0.133(11)(14) \,.
\end{equation}
Combining the two, the result for the spin of the nucleon is determined to be
\begin{equation}
J_N = \vec{J}_q + \vec{J}_g    =  0.541(62)(49) \,. 
\end{equation}
Within errors, the ETMC result is consistent with the proton spin being $1/2$. 

\section{Comparing results for quark angular momentum using Ji versus Jaffe-Manohar decompositions}
\label{sec:Jaffe}

\begin{figure}[tpb] 
\centerline{
\raisebox{20pt}{\includegraphics[width=0.42\linewidth]{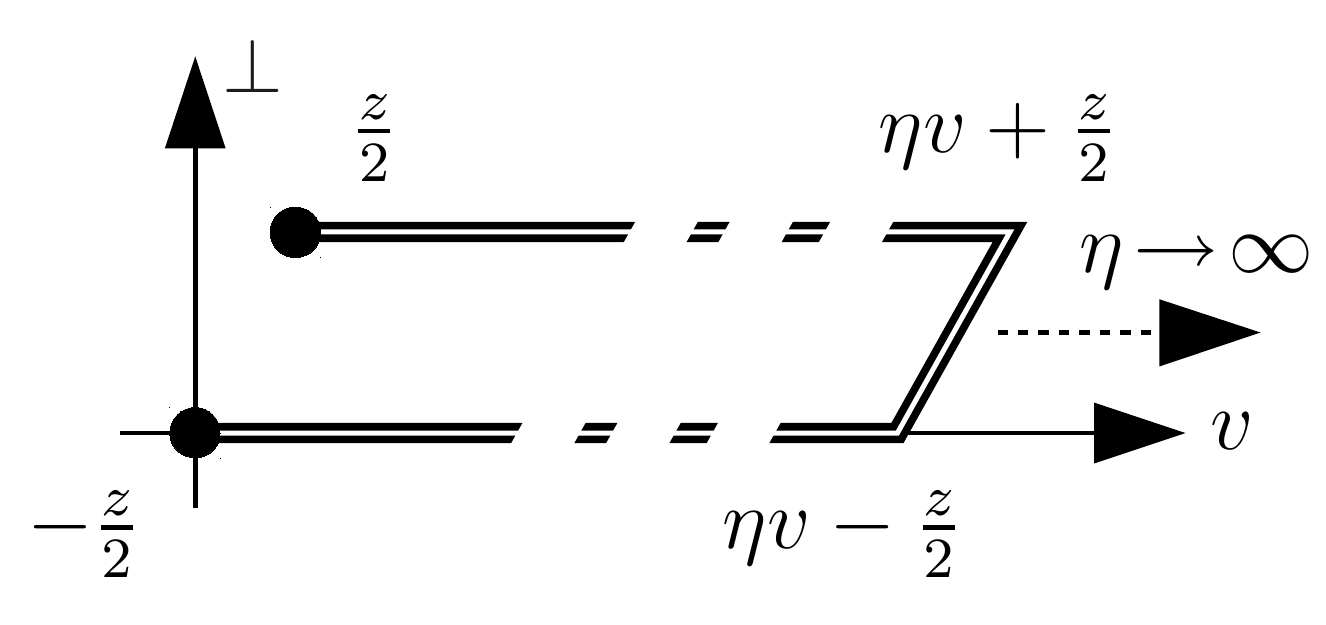}}\hspace{1cm}
\includegraphics[width=0.42\linewidth,viewport=0 195 612 597,clip=true]{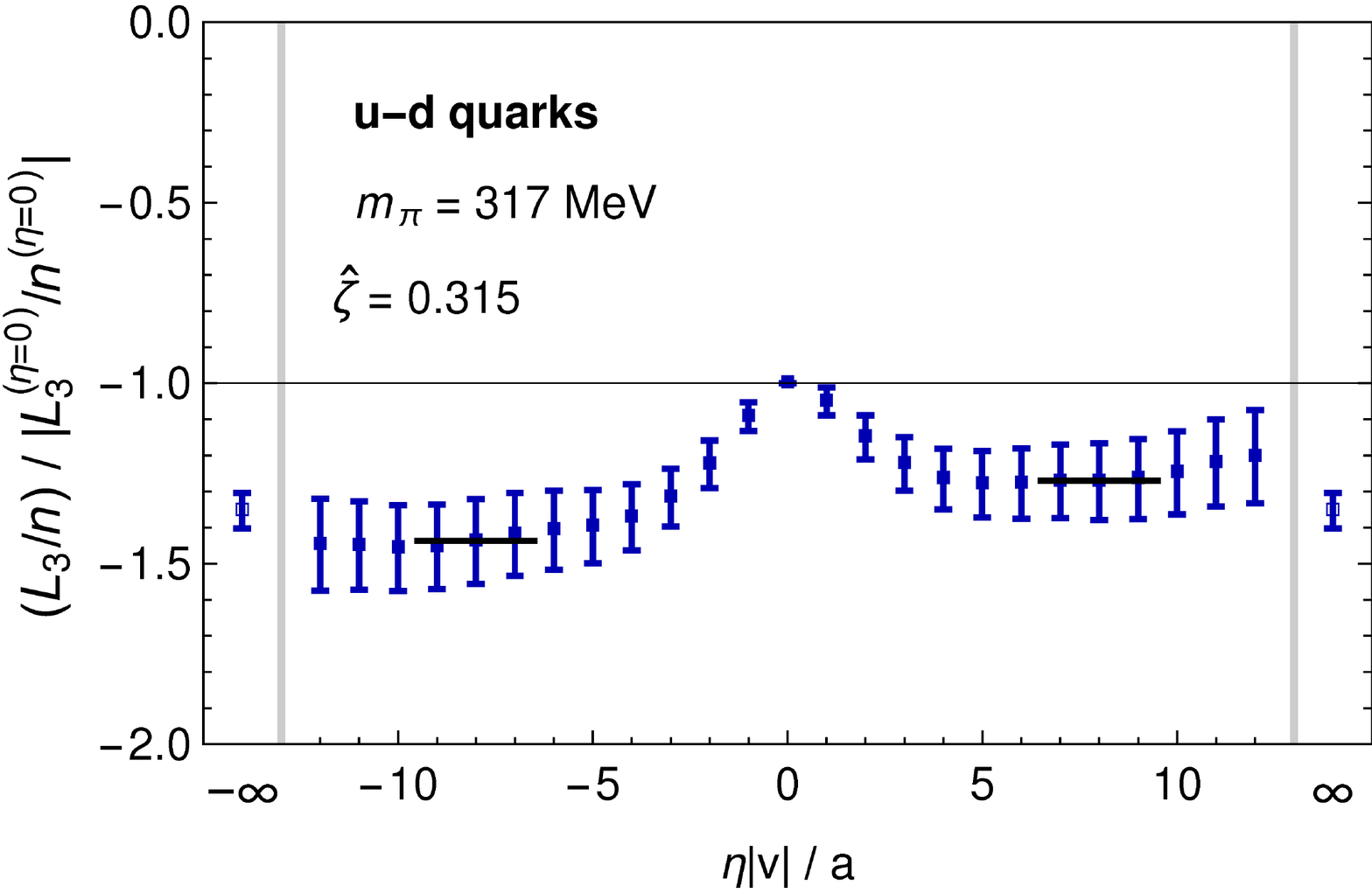}
}
\caption{(Left) The geometry of the staple shaped gauge connection between the 
quark and antiquark used to study the Ji and Jaffe-Manohar orbital angular momentum of the quarks. 
(Right) The ratio $|\vec{L}_{q}^{\rm JM}| /                                                                               
|\vec{L}_{q}^{\rm Ji}|$ obtained on a 2+1-flavor clover ensemble with $M_\pi = 317$~Mev 
and $a=0.114$~fm~\protect~\cite{Engelhardt:2019lyy}.  }
\label{fig:staple}
\end{figure}

Engelhardt~\cite{Engelhardt:2017miy,Engelhardt:2019lyy} has been
developing methods to directly calculate the orbital angular momentum (OAM) of
the quarks. The definition of OAM by Ji, 
\begin{equation}
\vec{L}_{q}^{\rm Ji}  =  \int d^3x  \ {q^\dag} \left[\vec x \times i \vec{D} \right] q \,,
\end{equation}
differs from that  defined on the light-cone by Jaffe-Manohar~\cite{Jaffe:1989jz},
\begin{equation}
\vec{L}_{q}^{\rm JM}  =  \int d^3x  \ {q^\dag} \left[\vec x \times i \vec{\nabla} \right] q \,,
\end{equation}
in the form of the spatial derivative term. The relevant matrix
elements required are of non-local quark bilinear operators connected
by a staple shaped gauge connection shown in Fig.~\ref{fig:staple}
(left). In this setup, the quark-antiquark is separated by distance
$z$ in a direction that is transverse to both the average nucleon
momentum $P$ and the momentum transfer $\Delta_T$; $p^\prime = P + \Delta_T$ and $p = P -
\Delta_T$; and the nucleon spin and the staple direction $v$ are taken
along the direction of $P$, which is typically taken to be the ``3''
direction. The matrix element of the operator with a straight link
path ($\eta=0$) gives $\vec{L}_{q}^{\rm Ji}$, while the Jaffe-Manohar
OAM, $\vec{L}_{q}^{\rm JM}$, is obtained in the limit $\eta \to
\infty$. First results for both are presented in Ref.~\cite{Engelhardt:2019lyy}. 

Results for the ratio $|\vec{L}_{q}^{\rm JM}| /
|\vec{L}_{q}^{\rm Ji}|$ are shown in Fig.~\ref{fig:staple}
(right). They indicate that $|\vec{L}_{q}^{\rm JM}| $ is about 40\%
larger than $ |\vec{L}_{q}^{\rm Ji}|$. The difference is interpreted
as the extra torque, due to final state interactions, accumulated by
the struck quark as it flies out of the proton. Following up on this
encouraging result, further work is in progress.

\section{Conclusions}
\label{sec:conclusions}

This review makes the case that calculations of the nucleon spin from
first principle simulations of lattice QCD are beginning to provide
results with control over all systematics.  Of the three contributions
analyzed, the best determined is the quark contribution
$\sum_{q=u,d,s,c} (\frac{1}{2} {\Delta q})$, followed by $J_q$ and
finally $J_g$ and the orbital angular momentum of the quarks. The
first results discussed are already consistent with phenomenology. The
PNMDE collaboration have presented results for $\sum_{q=u,d,s}
(\frac{1}{2} {\Delta q}$ with control over the various
systematics. They find $\sum_{q=u,d,s} (\frac{1}{2} {\Delta q} ) =
0.143(31)(36) $, consistent with the COMPASS value $0.13 < \frac{1}{2}
\Delta \Sigma < 0.18$ obtained at
3~GeV${}^2$~\cite{Adolph:2015saz}. At the same time, the PNDME
analysis makes a compelling case for the need for a new level of
control over all systematic uncertainties in order to obtain results
with $\le 10\%$ total error.

The ETMC collaboration has presented first results for $J_q$ and
$J_g$, and Engelhardt\cite{Engelhardt:2019lyy} for the orbital angular
momentum of quarks. Over the next five years, with anticipated
increase in computing resources, high precision results for all three
will become available and provide an accurate picture of their
relative contributions to the nucleon spin.

\acknowledgments
I thank the organizers of Spin 2018 for inviting me to give this
review and Prof. Lenisa for his hospitality.  On behalf of the PNDME
collaboration, I thank the MILC collaboration for sharing the
$2+1+1$-flavor HISQ ensembles generated by them and gratefully
acknowledge the computing facilities at, and resources provided by, 
NERSC, Oak Ridge OLCF, USQCD and LANL Institutional Computing.

\bibliographystyle{JHEP}
\bibliography{ref} 


\end{document}